%% file: main.tex
\documentclass[12pt]{iopart}
\usepackage{graphicx}
\usepackage{bm}
\usepackage{color}

\def\l{\langle}
\def\r{\rangle}

\begin{document}

\title{Ising model on the aperiodic Smith hat}

\author{Yutaka Okabe$^1$, Komajiro Niizeki$^{2}$\footnote{Professor emeritus} and Yoshiaki Araki$^3$}

\address{$^1$Department of Physics, Tokyo Metropolitan University, Hachioji, Tokyo 192-0397, Japan\\
$^2$Graduate School of Science, Tohoku University, Aoba-ku, Sendai 980-8578, Japan\\$^3$Japan Tessellation Design Association, Musashino-shi, Tokyo 180-0003, Japan}
\ead{okabe@phys.se.tmu.ac.jp}
\vspace{10pt}
\begin{indented}
  \item[]\today
\end{indented}
\begin{abstract}
Smith {\it et al} discovered an aperiodic monotile of 13-sided shape 
in 2023. It is called the `Smith hat' and 
consists of 8 kites. We deal with the statistical physics of the lattice 
of the kites, which we call the `Smith-kite lattice'. 
We studied the Ising model on the aperiodic Smith-kite lattice and 
the dual Smith-kite lattice using Monte Carlo simulations. 
We combined the Swendsen-Wang multi-cluster algorithm and 
the replica exchange method. 
We simulated systems up to the total spin number $939201$.  
Using the finite-size scaling analysis, 
we estimated the critical temperature on the Smith-kite lattice as 
$T_c/J=2.405 \pm 0.0005$ and that of the dual Smith-kite lattice 
as $T^{*}_{c}/J=2.143 \pm 0.0005$. 
Moreover, we confirmed the duality relation between 
the critical temperatures on the dual pair of aperiodic lattices, 
$\sinh(2J/T_c) \sinh(2J/T^{*}_{c}) = 1.000 \pm 0.001$. 
We also checked the duality relation for the nearest-neighbor correlation 
at the critical temperature, essentially the energy, 
$\epsilon(T_c)/\coth(2J/T_c) + \epsilon(T^{*}_c)/\coth(2J/T^{*}_c) 
= 1.000 \pm 0.001$.
\end{abstract}
\vspace{10pt}
\begin{indented}
  \item[]Keywords: Smith hat, aperiodic lattice, Ising model, Monte Carlo simulation, duality relation
\end{indented}

\maketitle

\section{Introduction}

It is possible to fill a two-dimensional plane with a single shape using regular polygons: equilateral triangles, squares, and regular hexagons.
It has long been a matter of interest to determine whether 
there are other shapes. 
Penrose~\cite{Penrose} showed that a plane can be filled with two shapes, 
a fat rhombus and a thin rhombus, where the golden ratio, 
an irrational number, is involved. 
The tilings obtained from these shapes do not have the usual periodic structure 
but creates a quasiperiodic structure. 
It exhibits the 5-fold (10-fold) symmetry. 
The quasicrystal \cite{Levine} was studied by experimental studies, 
and Shechtman {\it et al} discovered the 5-fold (10-fold) 
symmetry~\cite{Shechtman84}.  
The Nobel Prize in Chemistry 2011 was awarded to Shechtman 
``for the discovery of quasicrystals". 

Smith, Myers, Kaplan and Goodman-Strauss discovered an aperiodic monotile 
in 2023~\cite{Smith}. 
The shape is 13-sided and called the `hat'. 
It is able to fill an infinite plane without overlaps or gaps 
in a pattern that not only never repeats but also never can be repeated. 
Various studies have been conducted following 
this discovery~\cite{Smith2,Socolar,Baake,Akiyama}. 
In \cite{Socolar} and \cite{Baake}, the vertex set of the monotile 
is shown to be a two-dimensional quasiperiodic lattice (2DQL) 
which is obtained from a six-dimensional or a four-dimensional periodic lattice 
by the cut-and-projection method. 2DQLs are divided into two groups 
according to the nature of 
the projection window used in the cut-and-projection method. 
The window of a 2DQL belonging to one of the two groups 
is a polygon but the one belonging to the other group 
is a domain with a fractal boundary. 
A representative 2DQL of the first group is Penrose tiling P1, 
whose major tile is a regular pentagon. The monotile is shown, 
however, to belong to the second group~\cite{Socolar,Baake}. 

The material science of quasilattices has been studied experimentally 
and theoretically in various ways. Recent studies have focused 
on superconductivity~\cite{Deguchi,Kamiya}, electric structure~\cite{Fuchs} 
and magnetic properies~\cite{Nawa}. 
There has been much interest in the phase transitions of spin systems 
on quasicrystals compared to those of ordered and random systems.
Okabe and Niizeki~\cite{Okabe88_p} studied spin systems 
on a Penrose lattice mainly by means of Monte Carlo simulation. 
They showed that the Ising model on the Penrose lattice 
belongs to the same universality class as that 
on the two-dimensional (2D) regular lattice. 
In addition, they studied the dual lattice~\cite{Okabe88_dp}
and discussed the duality relation. 
A high-precision study was performed recently up to $N = 20,633,239$ 
using a GPU-based calculation~\cite{komura_Penrose}; 
A significant improvement in accuracy was obtained.

In contrast, no one has investigated up to present spin systems on a 2DQL 
belonging to one of the second group in the classification above. 
In this paper, we study the Ising models on the Smith-kite lattice 
and its dual lattice by means of the Monte Carlo simulation. 
We use the Swendsen-Wang (SW) multi-cluster spin-flip 
algorithm~\cite{sw87} combined 
with the replica exchange method of parallel tempering~\cite{Hukushima}. 
We organize the rest of the paper as follows. 
In Sec.~2, we explain the model and numerical method. 
We give the results in Sec.~3.  
We report the estimate of the critical temperature, and 
discuss the duality relation. 
The final section is devoted to 
the summary and discussions.

\section{Model and Numerical Method}

\subsection{Lattice}
\begin{figure}[h]
\begin{center}
\includegraphics[width=7.0cm]{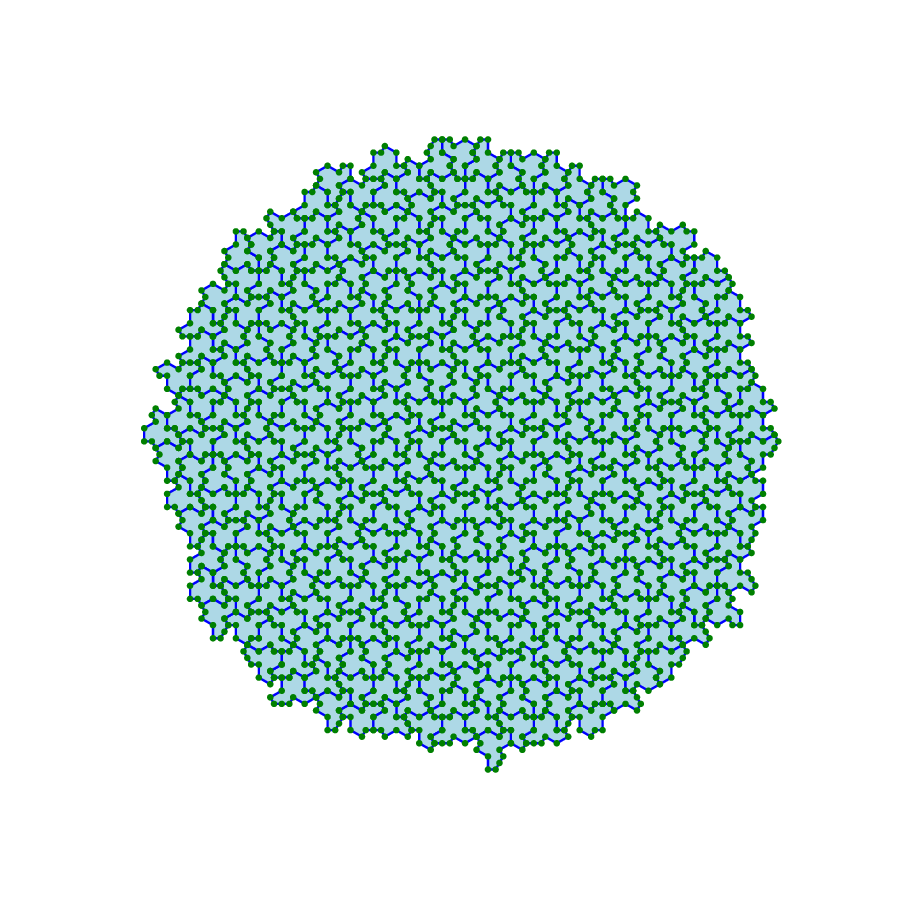}
\hspace*{4mm}
\includegraphics[width=7.0cm]{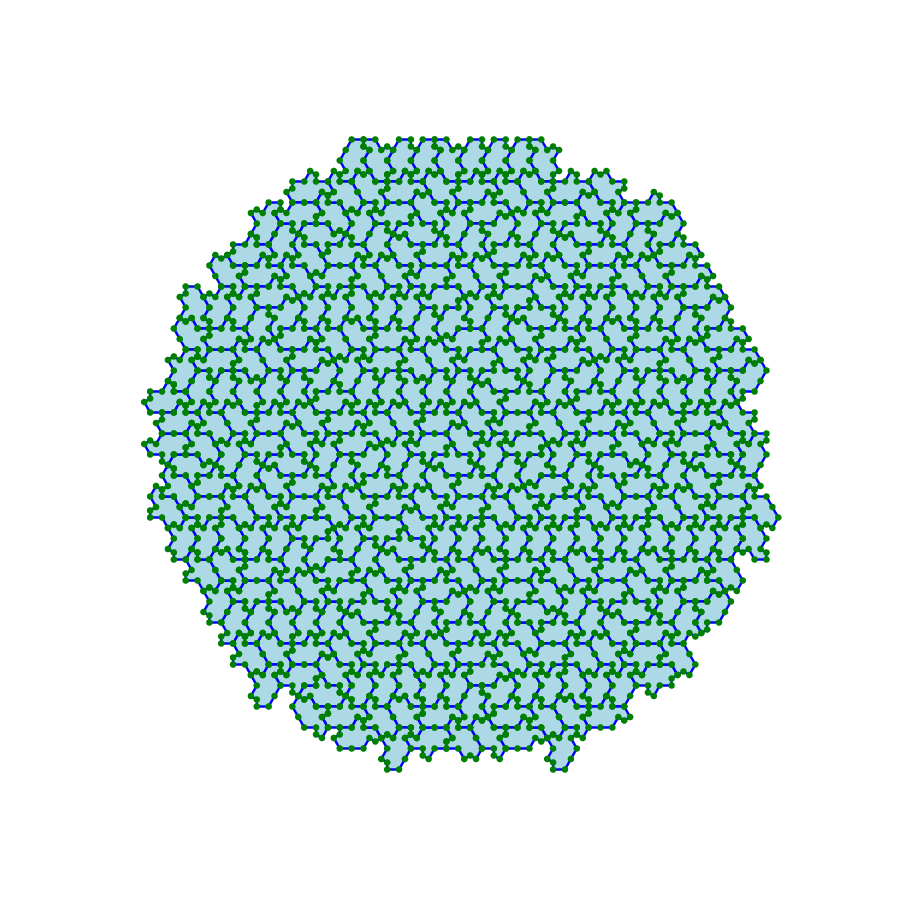}
\caption{
Smith-hat tiling of the ``small" size ($N_{\rm hat}=357$). 
We use the code released by Cheritat~\cite{Cheritat}. 
The left illusration is Tile($1,\sqrt{3}$) of 8 kites, 
whereas the right illusration is Tile($\sqrt{3},1$) of 10 kites.
}
\label{Smith_hat}
\end{center}
\end{figure}

Smith, Myers, Kaplan and Goodman-Strauss announced 
the discovery of an aperiodic monotile in March 2023~\cite{Smith}.
This brand new 13-sided shape is called `the Smith hat'. 
We here study statistical physics of spin systems on the Smith-hat tiling. 
As for the coordinate data, we use the code released 
by Cheritat~\cite{Cheritat}, 
which is a web application written in JavaScript. 
The 4-dimensional coordinate data of the hat vertices are given there, 
and by projecting them into two dimensions, 2-dimensional coordinate data 
can be obtained. 
Cheritat provides four sizes of Smith-hat tiling data. The number of hats 
is 357, 2490, 17077 and 117051 for ``small", ''medium", ``big" and 
``bigger" hat tiling, respectively. 
The Smith-hat tiling of the ``small" size is illustrated 
in figure~\ref{Smith_hat}.
It is noteworthy to mention a family of aperiodic monotiles~\cite{Smith}. 
The notation Tile($a,b$), where $a$ and $b$ are real numbers, 
is used to specify a shape of hat tile. 
Tile($1,\sqrt{3}$), which is the most popular, has 8 kites in the hat, 
whereas Tile($\sqrt{3},1$) has 10 kites in the hat. 
Although we mainly treat Tile($1,\sqrt{3}$) of a 13-sided monotile 
with 8 kites in this paper, we show both Tile($1,\sqrt{3}$) 
and Tile($\sqrt{3},1$) in figure~\ref{Smith_hat}.
In the site of Cheritat~\cite{Cheritat}, the default value of $\alpha_1$ 
is provided as 60 for Tile($1,\sqrt{3}$). When we treat Tile($\sqrt{3},1$), 
we set $\alpha_1$ as 30.

\begin{figure}[h]
\begin{center}
\includegraphics[width=6.4cm]{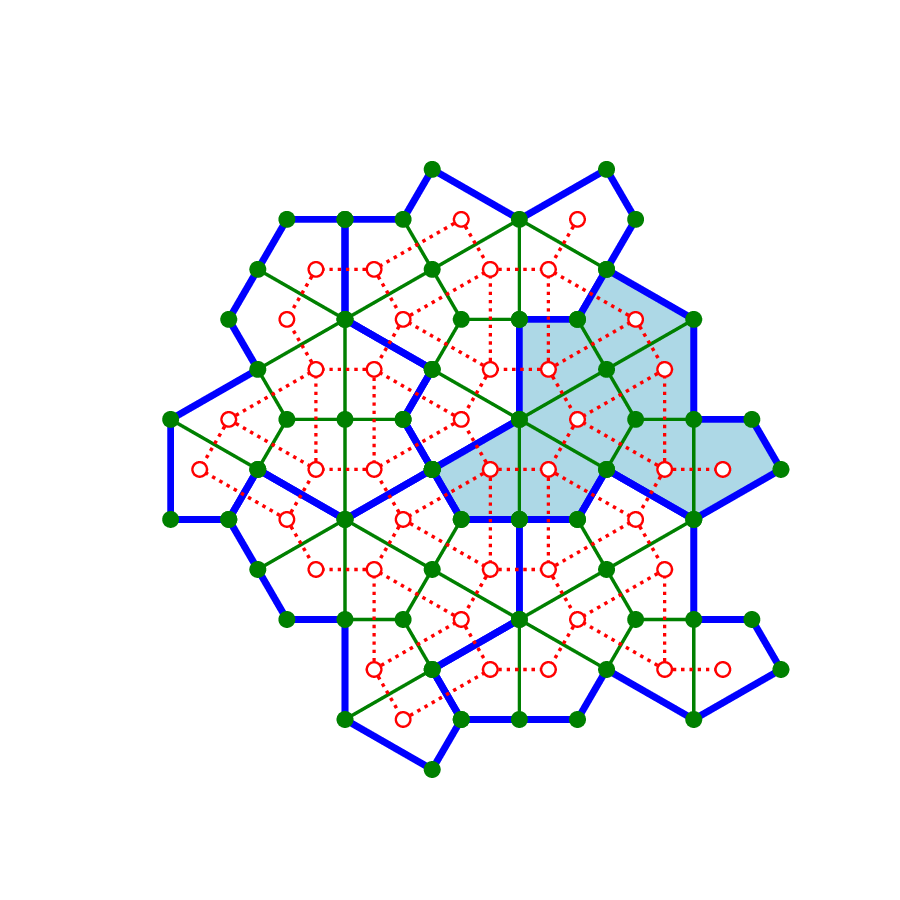}
\hspace*{8mm}
\includegraphics[width=6.8cm]{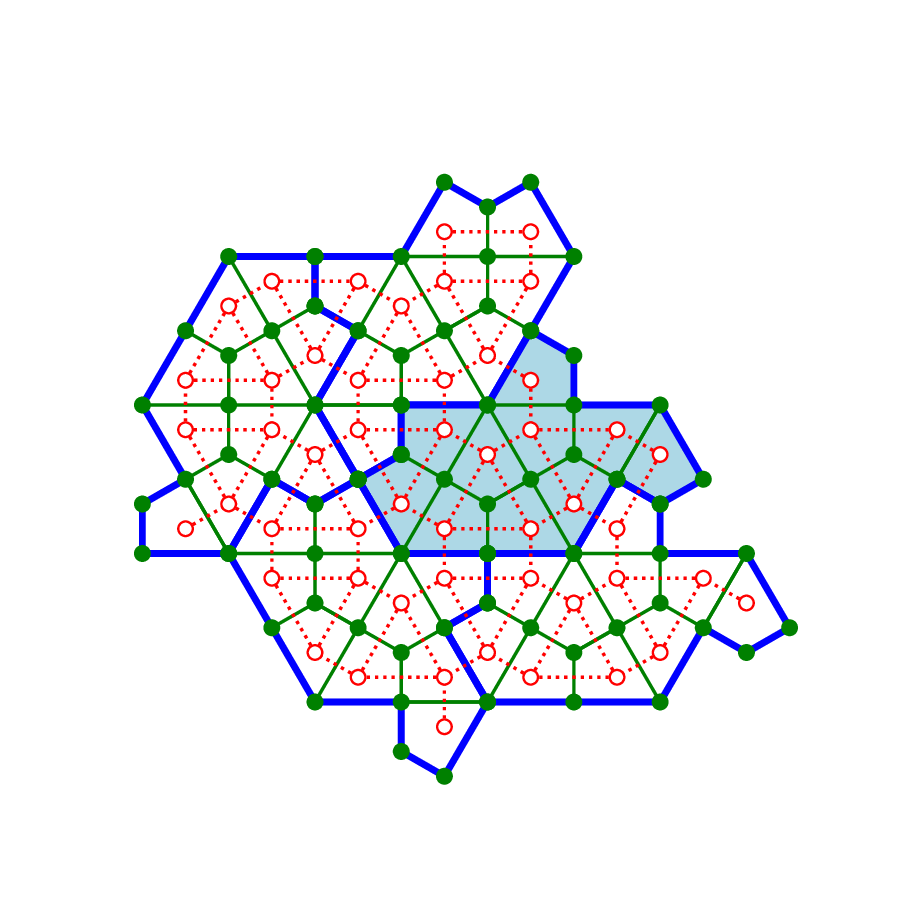}
\caption{
Smith-kite lattices. 
The left panel is Tile($1,\sqrt{3}$) of 8 kites, whereas 
the right panel is Tile($\sqrt{3},1$) of 10 kites. 
The edges of the Smith hat are denoted by blue lines. 
One hat is denoted by a shaded area. 
A hat consists of 8 or 10 kites, and the edges of 
the kites are denoted by thin green lines, 
which form the original Smith-kite lattice. 
The lattice points of the original Smith-kite lattice are denoted 
by green circles, whereas those of the dual Smith-kite lattice 
are denoted by empty red circles. 
We can see that there are extra 2 (4) lattice points in one 
8-kite (10-kite) hat.
The edges of the dual lattice are denoted by red dotted lines. 
In the Ising model, spins are put on the lattice points. 
}
\label{Smith_kite}
\end{center}
\end{figure}

We consider the lattice based on the kites. 
A hat consists of 8 or 10 kites. 
The vertices of the kites become the lattice points.  
There are extra 2 (4) lattice points in an 8-kite (10-kite) hat. 
We need the information on the connection of kite vertices; 
we make a table of connecting sites by checking the vertices shared 
by neighboring hats. 
One vertex is shared by two, three or four hats. 
In this way, we form a lattice made by 
the vertices of the kites, and we call it the original Smith-kite lattice. 
We also consider the dual lattice, which we call the dual Smith-kite lattice. 
The lattice points of the dual lattice are set in a kite. 
We make the connection between the dual lattice points 
by the two adjacent kites. 
These two lattices are dual, 
and we will examine the duality relations for dual lattices discussed 
by Kramers and Wannier~\cite{Kramers}. 
The original and dual Smith-kite lattices are illustrated 
in figure~\ref{Smith_kite}. 
The lattice points of the original lattice are denoted 
by green circles, whereas those of the dual lattice 
are denoted by empty red circles. 
The edges of the dual lattice are denoted by red dotted lines. 
We put spins on the lattice points when we study the Ising model. 

The sizes of the lattices are tabulated in table~\ref{table_hats}. 
This is the data for Tile($1,\sqrt{3}$) of 8 kites. 
The second column gives the number of hats for four sizes. 
The number of center of kites is 8 times the number of hats, 
whereas the number of vertex of kites is slightly larger 
because of the boundary. 
When we discuss the averaged values for the spins in the simulation, 
it is better to take an average over only inner sites, 
in order to reduce surface effects. Thus, we also list 
the number of inner spins. 

\begin{table}[h]
\caption{
Size of Smith-hat tiling provided by Cheritat~\cite{Cheritat}.
This is the data for Tile($1,\sqrt{3}$) of 8 kites. 
The original Smith-kite lattice is formed by the vertex of kites, 
whereas the dual Smith-kite lattice is formed by the center 
of kites.  The number of the inner sites will be used 
when taking an average in the simulation, and it will be 
denoted by $N$. 
}
\label{table_hats}
\begin{center}
\begin{tabular}{lrrrrrr}
\hline
\hline
& \multicolumn{1}{l}{hats} & \multicolumn{2}{l}{vertex of kites} 
& \multicolumn{2}{l}{center of kites}\\
& \ ~ & \ total & \ inner & \ total & \ inner \\
\hline
``small" & \ 357 & \ 3011 & \ 2807 & \ 2856 & \ 2655 \\
``medium" & \ 2490 & \ 20388 & \ 19794 & \ 19920 & \ 19372 \\
``big" & \ 17077 & \ 137686 & \ 136266 & \ 136616 & \ 135216 \\
``bigger" & \ 117051 & \ 939201 & \ 935529 & \ 936048 & \ 932721 \\
\hline
\end{tabular}
\end{center}
\end{table}

\subsection{Model and simulation}
We deal with the Hamiltonian of the Ising model given by
\begin{equation}
 \mathcal{H} = -J \sum_{\l i,j \r}s_{i}s_{j}, 
              \quad s_{i} = \pm 1,
\label{Ising}
\end{equation}
where $J$ is the coupling and $s_{i}$ is the spin at lattice site $i$. 
The lengths of the bonds in the kite are not all equal; 
since we are discussing the network structure of the lattice, 
we assume that the interaction constants all take constant $J$.
The summation is taken over nearest-neighbor pairs $\l i,j \r$.  
Here, we consider the case of the ferromagnetic coupling ($J>0$). 

In performing the Monte Carlo simulation, we employ the multi-cluster 
spin-flip SW algorithm~\cite{sw87} to escape from the problem 
of the long autocorrelation time near the critical temperature. 
We also combine the replica exchange method of 
parallel tempering~\cite{Hukushima}. 
We used the 20,000 Monte Carlo steps (MCS) after discarding 
the first 2,000 MCS for the ``small" and ``medium" sizes.  
We used the 10,000 MCS after the first 1,000 MCS for the ``big" size, 
and the 5,000 MCS after the 500 MCS for the ``bigger" size. 
Since the cluster flip algorithm~\cite{sw87} is used, 
the time required for equilibration is very short. 
Even for a largest system size, 100 steps are sufficient for equilibration. 
We discarded 1/10th of the measurement time steps. 
For temperature dependence, precise measurement is possible by applying 
the replica exchange method~\cite{Hukushima}. As the system size increases, 
it becomes necessary to take a narrower temperature difference 
to exchange temperatures. 
The detailed balance conditions are fulfilled for both algorithms
We made five independent runs 
with different random number sequences for each size 
to estimate numerical errors. In most plots the error bars are 
within the size of the marks. 
This study dealt with a sample of one configuration for each size. 
Because we are cutting finite sizes from an infinite system, 
we can take some other configurations. However, since we are 
dealing with sufficiently large sizes, we expect the dependence 
on configuration to be very small.

We make the Monte Carlo updates for the spins on all the sites.  
However, we take the average of physical quantities over only 
the spins on inner sites to reduce surface effects.
Surface effects are one of the important topics in the study 
of critical phenomena, and numerical investigation of surface effects 
is often performed based on the difference between periodic 
and free boundary conditions~\cite{Binder}. In this study, 
the bulk properties are the main subject, and we will 
consider the quantity excluding the surface layer.
The number of the inner sites for both original and dual lattices 
are tabulated in table~\ref{table_hats}, and 
in the following, we denote $N$ for the number of inner sites. 

\section{Results}
\subsection{Results of magnetization}
\begin{figure}[h]
\begin{center}
\includegraphics[width=7.5cm]{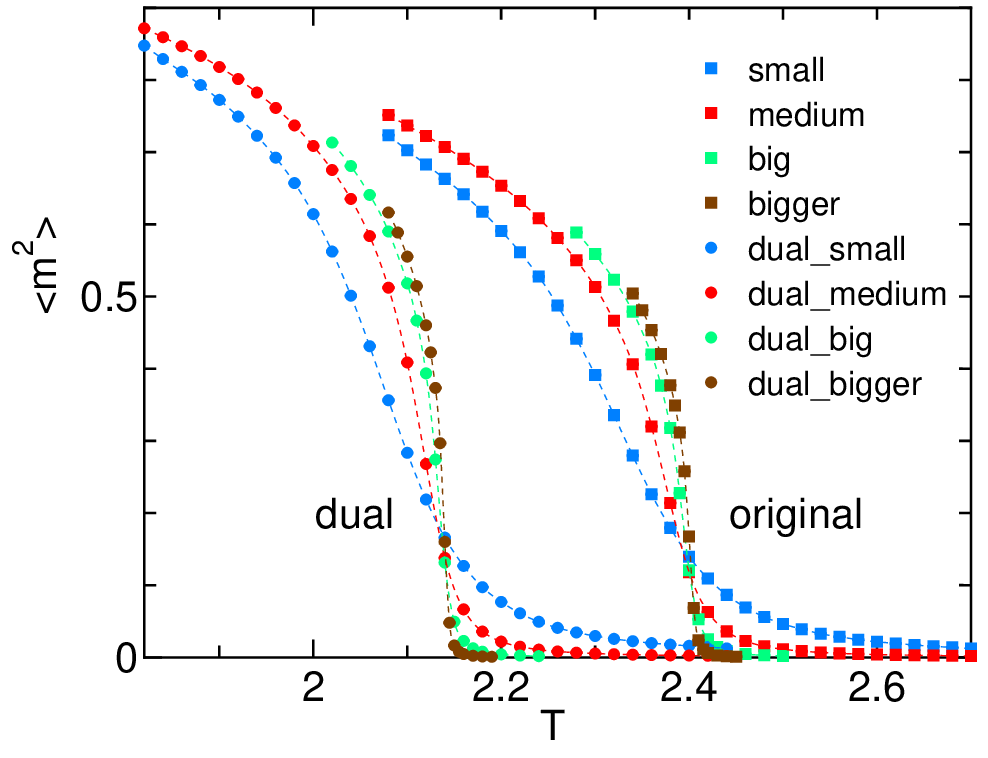}
\caption{
Squared magnetization of the Ising model on the Smith-kite lattices 
for both original and dual lattices. 
}
\label{m2}
\end{center}
\end{figure}

We present the simulation results of Tile($1,\sqrt{3}$) of 8 kites.
We first show the results of the magnetization per spin, $m = (1/N)\sum_i M_i$.
We plot the temperature dependence of the squared magnetization, $\l m^2 \r$, 
in figure~\ref{m2}. 
We show the data of the four sizes, ``small", ``medium", ``big", and ``bigger", 
for both original and dual lattices. 
The temperature is measured in units of $J$, that is, $J=1$. 
In figure~\ref{m2}, we observe the spontaneous magnetization for lower temperature.

\begin{figure}[h]
\begin{center}
\includegraphics[width=7.5cm]{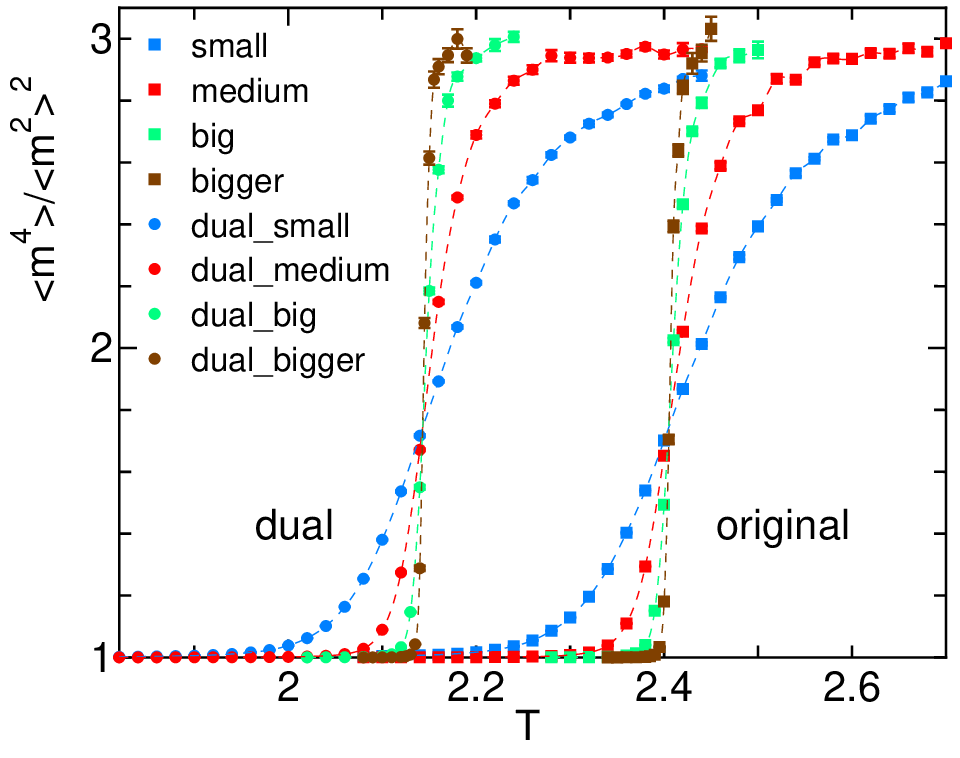}
\caption{
Moment ratio of the Ising model on the Smith-kite lattices 
for both original and dual lattices. 
}
\label{ratio}
\end{center}
\end{figure}

Binder~\cite{Binder81} pointed out that the moment ratio is effective 
for locating the critical temperature based 
on the finite-size scaling (FSS) analysis. 
We plot the temperature dependence of the moment ratio 
$\l m^4 \r/\l m^2 \r^2$ in figure~\ref{ratio}. 
We give the data of four sizes for both original and dual lattices.
This value approaches 1 for the ordered state ($T \to 0$), and 
approaches 3 for the disordered state ($T \to \infty$) 
because of Gaussian fluctuations.

The correlation length $\xi$ divereges as the temperature approaches 
the critical temperature as 
\begin{equation}
  \xi(T) \propto |T-T_c|^{-\nu}
\end{equation}
with the correlation-length exponent $\nu$. 
The FSS \cite{Barber,Cardy} postulates that 
near the critical temperature the physical quantities 
of finite systems with the linear size $L$ are 
scaled by $\xi/L$. 
The magnetization scales as 
\begin{equation}
  \l m^2 \r = L^{-2\beta/\nu} f(tL^{1/\nu}), 
\label{Eq_FSS_mag}
\end{equation}
where $\beta$ is the spontaneous-magnetization exponent 
and $t=T-T_c$. 
Then, the moment ratio becomes a single scaling variable such that
\begin{equation}
  \l m^4 \r/\l m^2 \r^2 = \tilde f(tL^{1/\nu}).
\end{equation}
It means that the plots $\l m^4 \r/\l m^2 \r^2$ of various sizes 
intersect at a single point, which leads to the estimate of 
the critical temperature.  It is to be noted that there 
are corrections to FSS for small sizes.
We see the crossing of data of different sizes in figure~\ref{ratio}. 
We here make a comment on the error bars.  At very high temperatures, 
this ratio becomes a 0/0 quantity, so the error bars may become large. 
However, just above the critical point, the error is still small 
even for the largest system size.

\begin{figure}[h]
\begin{center}
\includegraphics[width=13.0cm]{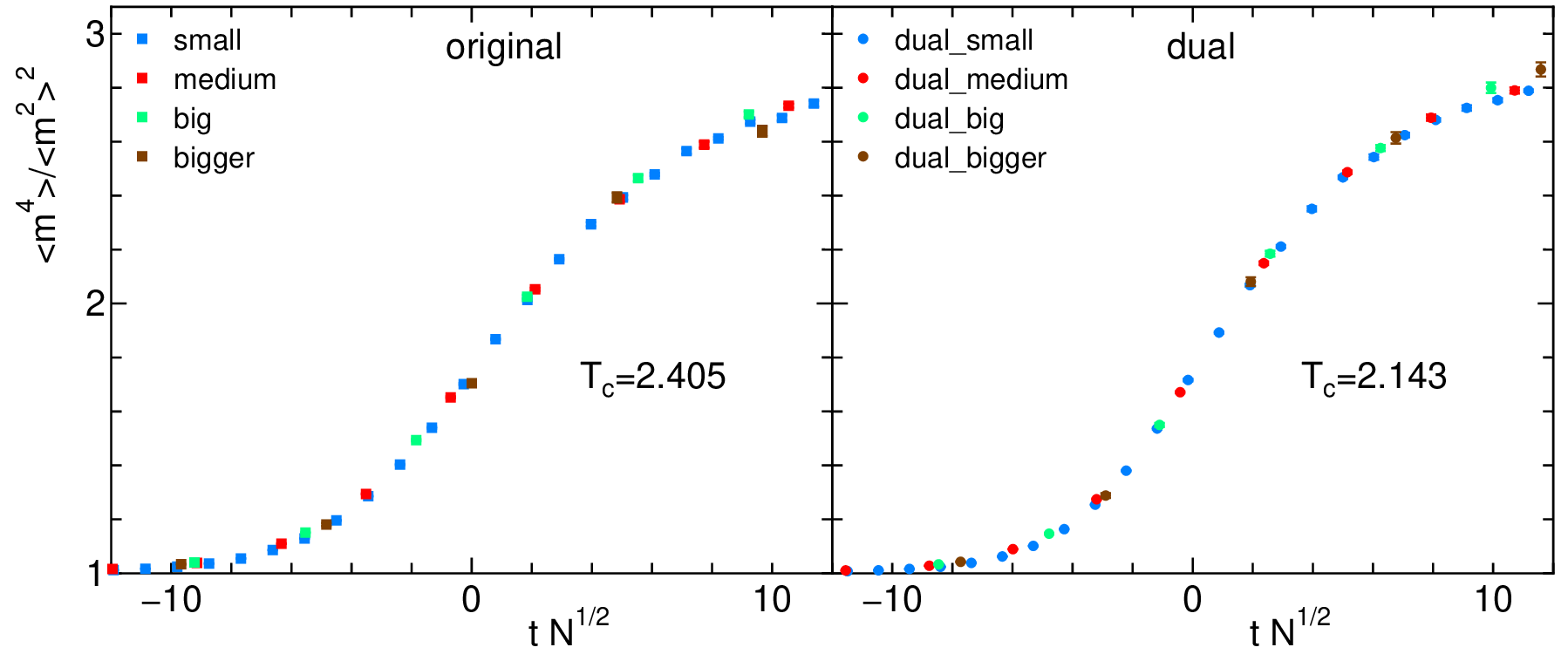}
\caption{
FSS of moment ratio for the Ising model 
on the Smith-kite lattices; $t=T-T_c$ and $\nu=1$. 
The left panel is the FSS plot for the 
original lattice, where we assume $T_c=2.405$, whereas the right panel 
is that for the dual lattice, where we assume $T^*_c=2.143$.}
\label{FSS_ratio}
\end{center}
\end{figure}

Assuming the critical temperature $T_c$, we try the FSS plots for 
both original and dual lattices in figure~\ref{FSS_ratio}. 
The linear size $L$ is $N^{1/2}$, and we use $\nu=1$ of the 
2D Ising value of the regular lattice. 
The universality of the critical exponents for quasiperiodic spin systems 
has been confirmed, and even if we analyze the critical exponent as a fitting parameter, we will obtain almost the same value as the exact solution 
using the FSS analysis. The left panel is the FSS plot for the 
original lattice, where we assume $T_c=2.405$, whereas the right panel 
is that for the dual lattice, where we assume $T^*_c=2.143$.

\begin{figure}[h]
\begin{center}
\includegraphics[width=13.0cm]{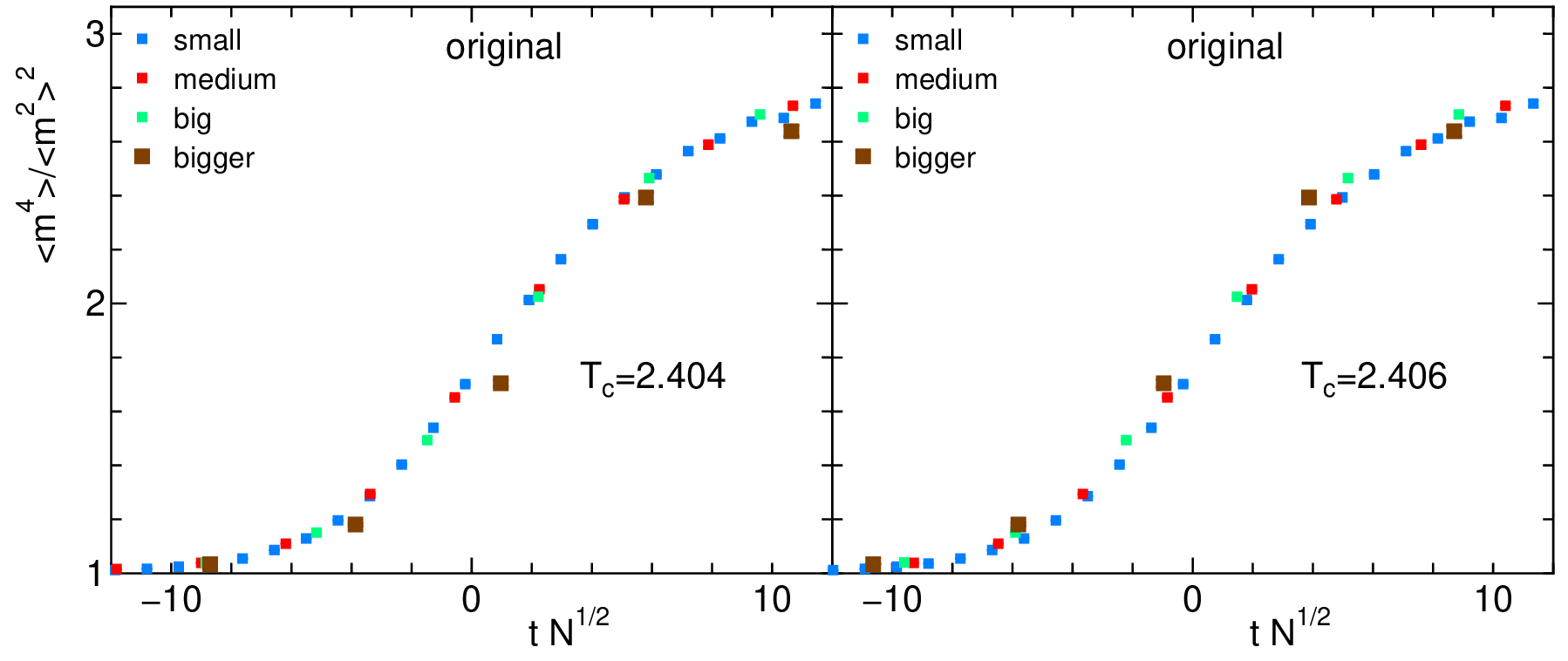}
\caption{
FSS trial of moment ratio for the Ising model on the original 
Smith-kite lattices. 
Choose $T_c=2.404$ (left panel) and $T_c=2.406$ (right panel). 
For clarity, larger marks are used for the ``bigger" lattice.}
\label{compare}
\end{center}
\end{figure}

For comparison, in figure~\ref{compare} we show the FSS trial 
when $T_c$ is assumed to be 2.404 and 2.406 
in the case of the original lattice. 
For clarity, we use larger marks for the ``bigger" lattice.
If you compare figure~\ref{compare} with the left panel of figure~\ref{FSS_ratio},
we can see that $T_c=2.405$ is a good choice. 
The trial plots of $T_c=2.404$ and $T_c=2.406$ in figure~\ref{compare} 
are choices of data to highlight the differences. 
We can obtain an estimate of the error in the narrow interval.
Thus, our estimates of the critical temperatures are 
\begin{equation}
  {\rm original \ lattice} \quad T_c = 2.405 \pm 0.0005
\label{Eq_Tc1}
\end{equation}
and
\begin{equation}
  {\rm dual \ lattice} \quad T^*_c = 2.143 \pm 0.0005.
\label{Eq_Tc2}
\end{equation}
%

\begin{figure}[h]
\begin{center}
\includegraphics[width=13.0cm]{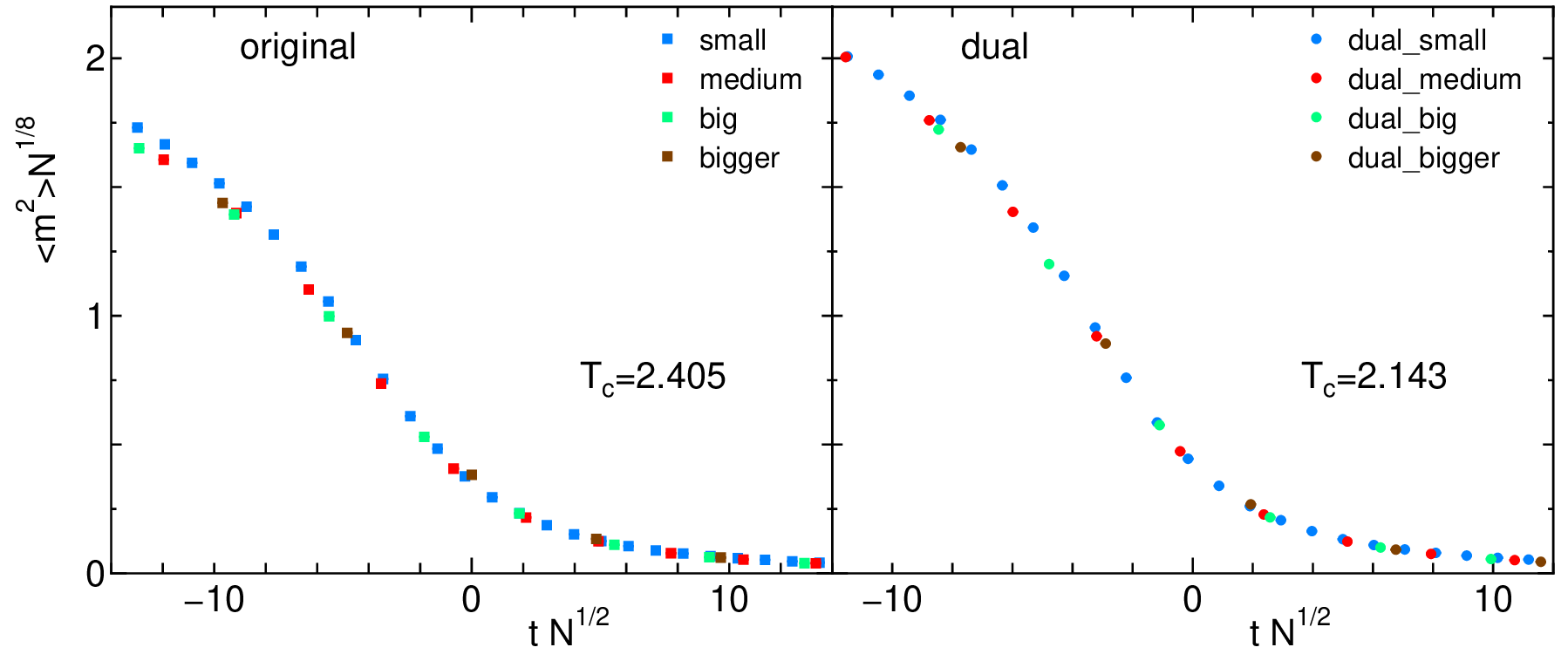}
\caption{
FSS of magnetization for the Ising model on the Smith-kite lattice; 
$t=T-T_c$, $\nu=1$ and $\beta=1/8$. 
The left panel is the FSS plot for the 
original lattice, where we assume $T_c=2.405$, whereas the right panel 
is that for the dual lattice, where we assume $T^*_c=2.143$.
}
\label{FSS_m2}
\end{center}
\end{figure}

Next, we show the FSS of magnetization, (\ref{Eq_FSS_mag}), 
in figure~\ref{FSS_m2}. Here, the critical exponents are those of 
the regular lattice, $\nu=1$ and $\beta=1/8$, and $T_c$ are chosen 
as the values of (\ref{Eq_Tc1}) and (\ref{Eq_Tc2}).
From figure~\ref{FSS_m2}, we can see that the FSS works quite well. 

We now consider the duality relation. 
A simple relation is known between the partition functions 
of the Ising models on a regular lattice and 
its dual lattice \cite{Kramers,Syozi72}. 
The duality relation between the critical temperatures, 
$T_{c}$ and $T^{*}_{c}$, is given by 
\begin{equation}
 \sinh(2J/T_c) \sinh(2J/T^{*}_{c}) = 1 .
\label{dual}
\end{equation}
Substituting the estimates of the critical temperatures on the Smith-kite 
lattice (\ref{Eq_Tc1}), and its dual lattice, (\ref{Eq_Tc2}),
we obtain $\sinh(2/T_c) \sinh(2/T^{*}_{c}) = 1.000 \pm 0.001$. 
The duality relation is well satisfied for the aperiodic monotile lattice.

\subsection{Results of Energy}
Another check of the duality is the duality relation for the critical energy, 
\begin{equation}
  \frac{\epsilon(T_c)}{\coth(2J/T_c)} 
  + \frac{\epsilon(T^{*}_c)}{\coth(2J/T^{*}_c)} = 1,
\label{dual_energy}
\end{equation}
where $\epsilon(T_c)$ is the nearest neighbor spin correlation 
at the critical temperature.  This quantity is related to 
the energy per spin at the critical temperature 
by $\epsilon(T_c) = - E(T_c)/2N$. 

\begin{figure}[h]
\begin{center}
\includegraphics[width=7.5cm]{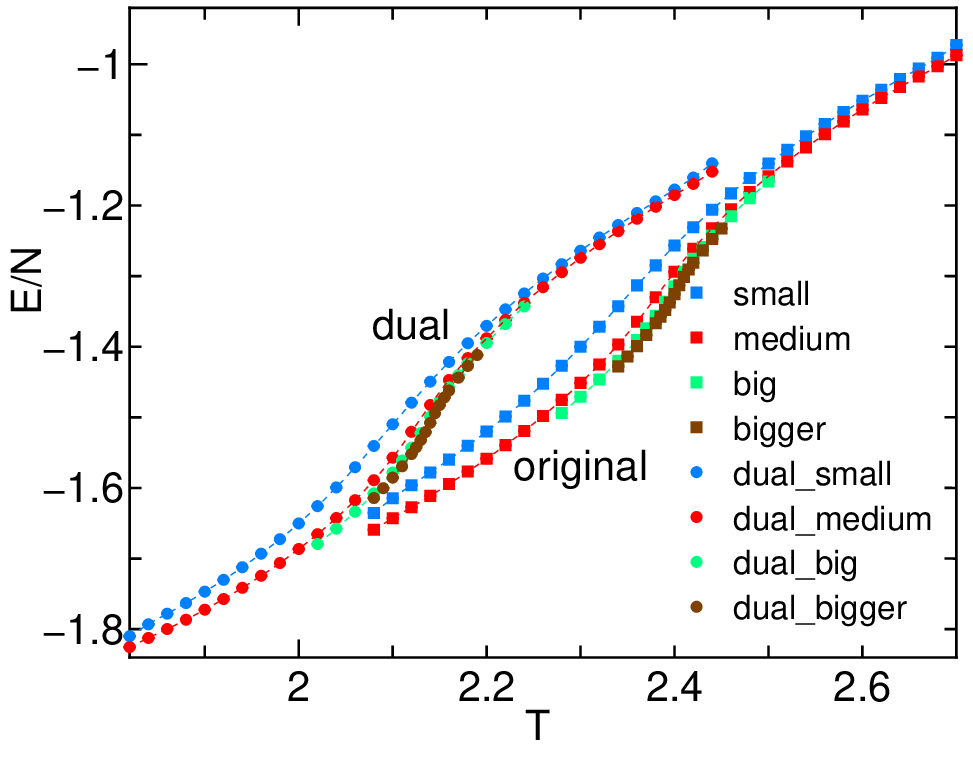}
\caption{
Energy of the Ising model on the Smith-kite lattices
for both original and dual lattices. 
}
\label{energy}
\end{center}
\end{figure}

\begin{figure}[h]
\begin{center}
\includegraphics[width=6.5cm]{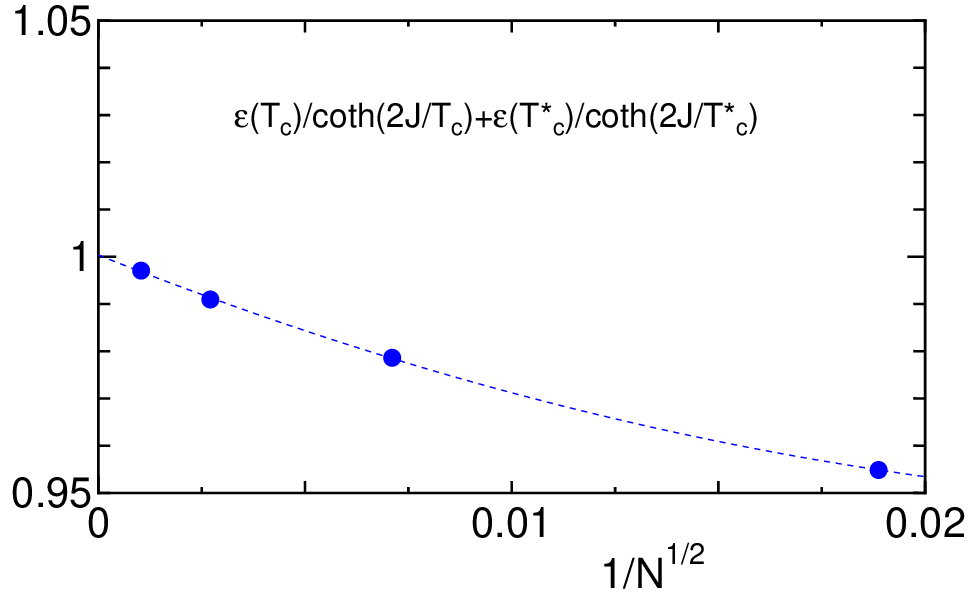}
\caption{
FSS of the critical energies of the Ising model 
on the Smith-kite lattices. We plot the left-hand side 
of (\ref{dual_energy}) as a function of $1/L=1/\sqrt{N}$. 
We also give the result of polynomial fitting 
by the dotted curve. 
}
\label{FSS_energy}
\end{center}
\end{figure}

The temperature dependence of the energy of the Ising model 
on the Smith-kite lattice and its dual lattice 
is shown in figure~\ref{energy}.
In order to estimate the energy at the critical temperature, 
we use the FSS for the critical energy, 
$(E(T_c,L) - E(T_c))/N \propto tL^{-(1-\alpha)/\nu}$. 
Here, $\alpha$ is the specific-heat exponent; it is 0 
for the 2D Ising model. 
The leading anomaly of the energy is $\alpha=0$, although 
the specific heat, the temperature derivative of the energy, 
has logarithmic divergence. 
Using the critical energies for each size, 
we plot the left-hand side of (\ref{dual_energy}), 
$ \epsilon(T_c)/\coth(2J/T_c) + \epsilon(T^{*}_c)/\coth(2J/T^{*}_c)$,
as a function of $1/L=1/\sqrt{N}$ in figure~\ref{FSS_energy}. 
The result of polynomial fitting (on the second order in powers of 
$1/\sqrt{N}$) is also shown by the dotted curve. 
Then we obtain the estimate of the left-hand side 
of (\ref{dual_energy}) as
$\epsilon(T_c)/\coth(2J/T_c)$
 + $\epsilon(T^{*}_c)/\coth(2J/T^{*}_c)$
 = $1.000 \pm 0.001$.
We also evaluated the critical energies by the same fitting procedure 
as $E(T_c)/N=-1.319$ for the original lattice and $E(T^*_c)/N=-1.505$ 
for the dual lattice.

The present result clearly indicates that 
the duality relations for the critical temperature and 
the critical energy hold for an aperiodic monotile lattice. 
In the case of the Penrose lattice, Komura and Okabe~\cite{komura_Penrose} 
presented the analytical argument for the duality relation 
based on the discussion by Kramers and Wannier \cite{Kramers}.  
They followed a detailed argument by Syozi \cite{Syozi72}.
The point is that the duality relation comes from the local dual transformation 
of the lattice, not from the regularity of the lattice. 
Thus, the same story can be applied to the present Smith-kite aperiodic lattice.

\subsection{Comparison with other lattices}

Here, we compare the present results of the critical temperature 
with those of other lattices. 
For the Ising models on the Penrose lattice and its dual lattices, 
the estimates of the critical temperatures by a high-precision study 
\cite{komura_Penrose} are
\begin{equation}
  T_c^{\rm Penrose} = 2.39781 \pm 0.00005
\end{equation}
and
\begin{equation}
  T_c^{\rm dual \ Penrose} = 2.14987 \pm 0.00005.
\end{equation}
In the simulation of the high-precision study~\cite{komura_Penrose}, 
the {\it periodic} Penrose lattice with a large unit cell was used. 
The formation of the Penrose tiling is associated with the golden ratio, 
which is the irrational number $\tau=(1+\sqrt{5})/2$. 
The {\it periodic} Penrose lattice is constructed by 
the continued-fraction expansion of the irrational number. 
The number of the lattice sites in the unit cell 
depend on the truncated level of the continued-fraction expansion.

It is also instructive to compare the present data 
with the data of regular lattices, where the coordination number 
is 4 on average. 
The critical temperatures of the Ising model on the kagome lattice 
and its dual lattice, the diced lattice, were exactly obtained
\cite{Syozi,Kano}. See also the calculation of the Fisher's zero for 
the kagome and diced lattices \cite{Feldmann}. 
The obtained exact critical temperatures are 
\begin{equation}
  T_c^{\rm kagome} = \frac{4}{\ln(3+2\sqrt{3})}=2.143319 \cdots
\end{equation}
and
\begin{equation}
  T_c^{\rm diced} = \frac{2}{\cosh^{-1}\frac{1+\sqrt{3}}{2}}
  =2.405457 \cdots.
\end{equation}
Of course, these critical temperatures satisfy the duality relation 
(\ref{dual}).

In table~\ref{Tc}, we compare the $T_c$'s of the Ising models 
on the (quasi-)lattices with the 4 coordination number. 
We also give the information on the lattice structure in table~\ref{Tc}. 
The square lattice is self-dual, and we added it for reference; 
\begin{equation}
  T_c^{\rm square} = \frac{2}{\ln(1+\sqrt{2})} = 2.26919 \cdots.
\end{equation}
The Smith-kite lattice, Penrose lattice, diced lattice are bipartite lattices, 
and take several values for the coordination number; 
there is no frustration for antiferromagnetic systems. 
On the other hand, the dual Smith-kite lattice, dual Penrose lattice, 
kagome lattice are not bipartite, and the coordination number is fixed as 4. 
Because there are ``odd-numbered" rings, there appears 
frustration for antiferromagnetic systems. 
The duality relations are satisfied for all the dual pairs. 
The $T_c$'s of the Smith-kite lattice, Penrose lattice, diced lattice are 
almost the same, which reflects upon the similarity of the lattices.

\begin{table}[h]
\caption{
Critical temperatures of Ising models on various (quasi-)lattices 
whose (average) coordination number is 4. 
We also give the information on the lattice structure in the table. 
The numbers in the parentheses are the average coordination number. 
}
\label{Tc}
\begin{center}
\begin{tabular}{lllll}
\hline
\hline
lattice & Smith-kite & d-Smith-kite & Penrose & d-Penrose\\
\hline
coordination number & 3,4,6 (4) & 4 & 3,4,5,6,7 (4) & 4\\
bipartite & yes & no & yes & no\\
frustration (AF) & no & yes & no & yes\\
$T_c/J$ & 2.405 & 2.143 & 2.39781 & 2.14987\\
\hline
lattice & diced & kagome & square & \\
\hline
coordination number & 3,6 (4) & 4 & 4 & \\
bipartite & yes & no & yes & \\
frustration (AF) & no & yes & no & \\
$T_c/J$ & 2.405457 $\cdots$ & 2.143319 $\cdots$ & 2.26919 $\cdots$ & \\
\hline
\end{tabular}
\end{center}
\end{table}

\section{Summary and discussions}

To summarize, we have studied the Ising models on the Smith-kite lattice 
and the dual Smith-kite lattice  by means of  the SW multi-cluster 
spin-flip algorithm.
We have estimated the critical temperatures on these lattices, 
and have shown that the finite-size scaling works quite well.  
The estimated critical temperature on the Smith-kite lattice  
is $T_c/J=2.405 \pm 0.0005$ and that on the dual Smith-kite lattice is 
$T^{*}_{c}/J=2.143 \pm 0.0005$. 
We have checked the duality relation. 
From the results of critical temperatures on the Smith-kite lattice 
and its dual lattice, we have obtained $\sinh(2J/T_c) \sinh(2J/T^{*}_{c}) 
= 1.000 \pm 0.001$. 
We have also shown the duality relation for the energy, 
$\epsilon(T_c)/\coth(2J/T_c)$
 + $\epsilon(T^{*}_c)/\coth(2J/T^{*}_c)$
 = $1.000 \pm 0.001$.
These numerical results clearly indicate 
that the duality relation is satisfied 
for an aperiodic monotile lattice.  
We have compared the present results with those of the Ising 
models on other lattices, including the Penrose lattice 
of quasicrystal and the regular lattices (kagome and dice lattices). 
The average coodination numbers of all these lattices are four.

We presented the results of Tile($1,\sqrt{3}$) with 8 kites. 
We also performed the simulation for Tile($\sqrt{3},1$) with 10 kites. 
The results obtained are in close numerical agreement. 
Although not in perfect agreement, the difference is within 0.05\% 
in the estimate of the critical temperatures. 
This can be attributed to the fact that both are tiling families 
of the same origin.

In the simulation of the Penrose lattice, the {\it periodic} lattice 
was used~\cite{Okabe88_p,Okabe88_dp,komura_Penrose}. 
For the Smith-kite lattice, it will be helpful 
to use {\it periodic} lattice for high-precision simulation.  
Similar procedure to approximate the irrational number 
by a series of rational numbers could be applied to the case 
of Smith-kite lattice.

From the viewpoint of statistical physics, 
we have considered the Smith-kite lattice consisting 
of aperiodic monotiles. The study of electric systems, for example, 
on this aperiodic lattice is desirable. 
Quite recently, the spectral and transport properties 
of a tight-binding model defined on the Hat~\cite{Schirmann} 
and the dimer model on the Hat~\cite{Singh} have been discussed.

Finally, we would like to comment on the study of tiling itself. 
The apeioric Smith tiling still has much to be studied 
from a mathematical science standpoint. 
The discovery of a new aperiodic monotile is also desirable.

\section*{Data availability statement}
All data that support the findings of this study are included within the article.

\section*{Acknowledgments}
We thank Arnaud Cheritat for allowing us to use the program code he created. 
This work was supported by JSPS KAKENHI Grant Number JP22K03472.

\section*{ORCID iDs}
Yutaka Okabe https://orcid.org/0000-0002-9434-5653


\include{reference}

\end{document}

%% file: reference.tex
\newcommand{\mybibitem}[6]{\bibitem{#1} #2 #6 #3 {\it #4} #5}